\documentclass[twocolumn,showpacs,showkeys,amsmath,amssymb,nofootinbib,prd]{revtex4}
\usepackage{graphicx}
\usepackage{times}

\begin{document}

\title{Dark matter annihilation near a black hole: plateau vs. weak cusp}
\author{Eugene Vasiliev}
\email{eugvas@lpi.ru}
\affiliation{Lebedev Physical Institute, Leninsky pr. 53, Moscow, Russia}

\date{August 1, 2007}

\begin{abstract}
Dark matter annihilation in so-called ``spikes'' near black holes is believed to be an important method 
of indirect dark matter detection. 
In the case of circular particle orbits, the density profile of dark matter has a plateau at small radii, 
the maximal density being limited by the annihilation cross-section. 
However, in the general case of arbitrary velocity anisotropy the situation is different. 
Particulary, for isotropic velocity distribution the density profile cannot be shallower than $r^{-1/2}$ 
in the very center. 
Indeed, a detailed study reveals that in many cases the term ``annihilation plateau'' is misleading, 
as the density actually continues to rise towards small radii and forms a weak cusp, 
$\rho \propto r^{-(\beta+1/2)}$, where $\beta$ is the anisotropy coefficient.
The annihilation flux, however, does not change much in the latter case, if averaged over an area larger 
than the annihilation radius.
\end{abstract}

\pacs{95.35.+d, 95.55.Ka}

\maketitle

\section{Introduction} 
One of the most promising methods of indirect dark matter search is the detection of annihilation signal from 
so-called dark matter ``spikes'' around a black hole (either a supermasive black hole in galactic center or 
an intermediate-mass black hole). The existence of spikes is predicted by adiabatic growth model 
\cite{a9906391}. If the initial density profile is cusped, $\rho\propto r^{-\gamma}$, then the adiabatically 
grown spike will also have power-law profile $\rho'\propto r^{-\gamma'}$, with $\gamma'=(9-2\gamma)/(4-\gamma)$.
If the initial profile is cored, then the spike will have $\rho'\propto r^{-3/2}$. 

It is evident that densities can reach very high values for small radii (but greater than $r_h=2GM_{bh}/c^2$). 
The annihilation rate of dark matter particles is therefore high enough to reduce the density. The usual 
argument is the following: consider the dark matter distribution function $f({\bf r}, {\bf v})$. The 
annihilation rate at a certain point is given by
\begin{equation}  \label{ann_rate_local}
\frac{\partial f({\bf r}, {\bf v})}{\partial t} = -\frac{\rho({\bf r})}{m_\chi} \langle \sigma v\rangle \,f \,.
\end{equation}
Here $m_\chi$ is the particle mass, $\langle \sigma v\rangle$ is the annihilation cross-section times 
relative velocity, which is independent of $v$. 

If particles have circular orbits and the density is spherically symmetric, then one may integrate $f$ over 
$d\bf v$ and obtain 
\begin{equation}  \label{dens_evolution_circ}
\frac{\partial \rho(r)}{\partial t} = -\frac{\rho^2}{\rho_a t}\;, \quad 
\rho(r) = \frac{\rho_0(r) \rho_a}{\rho_0+\rho_a} \;, \quad
\rho_a = \frac{m_\chi}{\langle \sigma v\rangle t} \;.
\end{equation}
Here $\rho_0 = \rho(r,t=0)$ is the initial density, and $\rho_a$ is called annihilation plateau density.
One can see that indeed for small radii the density approaches $\rho_a$, and the ``height'' of the plateau 
decreases in time.

In the general case of non-circular orbits (but still having spherical symmetry) it is easier to switch from 
\{{\bf r}, {\bf v}\} variables to \{$E, L$\} variables (energy and angular momentum per unit mass). Then 
Eq.(\ref{ann_rate_local}) becomes
\begin{equation}  \label{ann_rate_aver}
\frac{\partial f(E,L)}{\partial t} = -\frac{\tilde \rho}{\rho_a t}\,f \;,\quad
\tilde \rho = \frac{1}{T}\oint \rho(r) \frac{dr}{v_r} \;.
\end{equation}
$\tilde \rho$ is the orbit-averaged density. For convenience, we replace $L$ for $R=L^2/L_c^2$, where 
$L_c=GM_{bh}/\sqrt{-2E}$ is the angular momentum of a circular orbit (and hence $0\leqslant R \leqslant 1$). 
The radius of a circular orbit is $r_c = GM_{bh}/(-2E)$, and we substitute $r=xr_c$.
Then we have
\begin{equation}  \label{tilde_rho}
\tilde \rho = \rho(r_c) \int_{1-\sqrt{1-R}}^{1+\sqrt{1-R}} \frac{\rho(x r_c)}{\rho(r_c)} 
\frac{dx}{\pi \sqrt{2/x-1-R/x^2}} \;,
\end{equation}

Assume we have density profile $\rho(r)\propto r^{-\gamma}$, and a velocity anisotropy parameter 
$\beta = 1-{\sigma_t^2}/{2\sigma_r^2}$ \cite{BT}. The case $\beta=-\infty$ corresponds to circular 
orbits, $\beta=0$ is the isotropic case, $0<\beta<1$ is the case of radial velocity anisotropy. 

The caveat is that in the isotropic case the density profile cannot be shallower than $r^{-0.5}$ in the 
center (or, more generally, $\gamma \geqslant \beta+1/2$ \cite{a0511686, a0605070}). This means that we cannot get  
annihilation plateau with constant density for initially isotropic velocity distribution. One might argue 
that the annihilation introduces tangential velocity anisotropy to the degree compatible with flat density 
distribution, but in fact $\tilde \rho(R)$ is {\sl increasing} function of $R$ for $0<\gamma<1$, and the 
circular orbits are indeed depopulated. Thus, a careful examination is needed in the general case, as noted 
in \cite{a0703236}. This is the aim of the present study.

\section{A qualitative argument} 


A simple argument can explain the impossibility to create density profile more shallow than 
$r^{-(\beta+1/2)}$.

Consider the distribution function of the following broken power-law form: 
\begin{equation}  \label{DF_brokenPL}
f(E,R) = f_0 R^{-\beta} \left\{ \begin{array}{rcl} (E/E_0)^{p_1} &,& |E|>|E_0| \\ (E/E_0)^{p_2} &,& |E_0|>|E|>0 \,.
\end{array} \right.
\end{equation}
We assume that $p_1<p_2$, i.e. the distribution function is convex in $\log E-\log r$ coordinates.
It is a simple exercise to show that the velocity anisotropy parameter equals exactly $\beta$ in the above 
expression.

We now demonstrate that if $p_1>\beta-1$, then the density in the region $r\ll r_0=GM_{bh}/(-E_0)$ is 
determined by the distribution function at $|E|>|E_0|$, and $\rho \propto r^{-(p_1+3/2)}$ (steeper than 
$r^{-(\beta+1/2)}$). In the opposite case, however, the density in the region $r<r_0$ is determined by 
distribution function {\sl outside} that region (i.e. with $|E|<|E_0|$), provided that $p_2-\beta>-1$.

The density is given by expression
\begin{equation}  \label{density}
\begin{array}{rcl} 
\rho(r) &=& \displaystyle \sqrt{2}\pi \left(\frac{GM_{bh}}{r}\right)^{3/2} \int_0^1 \frac{d\varepsilon}{\varepsilon} \\
  && \displaystyle \times \int_0^{4\varepsilon(1-\varepsilon)} \frac{dR\; f(-\varepsilon\,GM_{bh}/r, R)}{\sqrt{1-\varepsilon-R/4\varepsilon}} \,.
\end{array}
\end{equation}
Here $\varepsilon=-E\,r/GM_{bh}$ is dimensionless energy. 

If we are interested in $r\ll r_0$, then the integral can be split in two constituents: 
$\rho_1=\int_{\varepsilon_0}^1\cdots$ and $\rho_2=\int_0^{\varepsilon_0}\cdots$, 
representing the contribution of inner and outer areas, correspondingly. 
$\varepsilon_0 = -E_0\,r/GM_{bh} \ll 1$ by condition.
Then we get
\begin{equation}  \label{density2}
\begin{array}{rcl} 
\rho_i &=& \displaystyle  f_0\:\mathrm B(1/2,1-\beta)\: 4^{1-\beta}\sqrt{2}\pi\, \left(\frac{GM_{bh}}{r}\right)^{3/2} \\
  &&\times \displaystyle \int d\varepsilon\: (1-\varepsilon)^{-\beta+1/2} \:
           \varepsilon^{p_i-\beta}\: \varepsilon_0^{-p_i}\;,\\
  i &=& 1,2\mbox{\Large \strut} \;,
\end{array}
\end{equation}
where limits of integration are given above.

Now there are two cases for $\rho_1$:
if $p_1-\beta>-1$, then the integrand is finite as $\varepsilon_0 \to 0$ and $\rho_1 \propto r^{-(3/2+p_1)}$.
In the opposite case the integral diverges as $\varepsilon_0^{p_1-\beta+1}$ and $\rho_1 \propto r^{-(1/2+\beta)}$.
The second integral, $\rho_2$, is always $\propto r^{-(1/2+\beta)}$.
Hence in the first case $\rho_1 \gg \rho_2$ and $\rho$ is determined by $p_1$, while in the second case they have 
the same dependence on $r$ and $\rho \propto r^{-(\beta+1/2)}$. 

What has this to do with annihilations? In the case $\beta<-1/2$ we may get density plateau in the center due 
to annihilations, which corresponds to $p_1=-3/2$. However, if initially $\beta\geqslant -1/2$, a constant 
density core cannot develop. Instead a sort of broken power-law density profile will emerge: 
\begin{equation}  \label{density_broken}
\rho \propto \begin{array}{rcl}  r^{-(\beta+1/2)} &,& r<r_0 \\ r^{-(p_2+3/2)} &,& r>r_0 \end{array}
\end{equation}
The break radius $r_0$ is the same as for density plateau, i.e. the density profile outside this radius 
is $\rho_a\: (r/r_0)^{-(p_2+3/2)}$. 
Particles outside $r_0$ are still not much affected by annihilation, since for them $\tilde\rho \sim \rho(r) < 
\rho_a$. Inside $r_0$, or, say another way, for $|E|>|E_0|$, the distribution function is depleded more 
rapidly than in the case of annihilation core, since the average density that particle ``feels'' is higher.
The boundary $r_0$ corresponds to the intermediate area where $\tilde\rho \approx \rho_a(t)$.

\section{Numerical calculations} 

\begin{figure}[t] 
$$\includegraphics{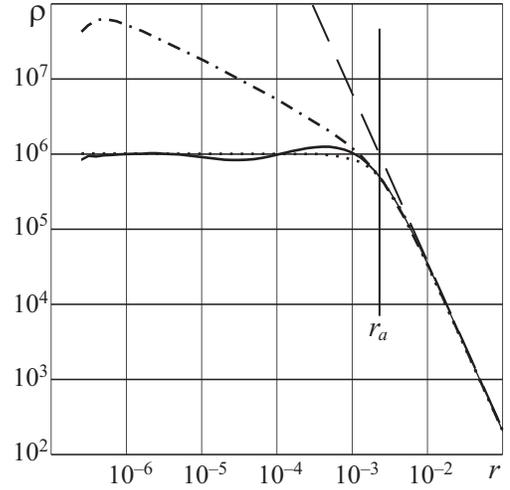} $$
\caption{
Density profiles after a certain time of evolution $t_1$ (see text) for different values of $\beta$: 
dotted line -- circular orbits ($\beta=-\infty$); solid line -- $\beta=-2$, dotted-dashed line -- $\beta=0$. 
In the latter case a weak cusp $\rho \propto r^{-1/2}$ develops instead of constant-density plateau. 
Initial density profile $\rho \propto r^{-9/4}$ is shown by the long-dashed line.
Radius and density are scaled to the black hole influence radius and corresponding density.
Annihilation plateau density is $\rho_a=10^6$, corresponding annihilation radius $r_a=2.2\cdot 10^{-3}$ is 
denoted by the vertical line.
} \label{fig_dens}
\end{figure} 

\begin{figure}[t] 
$$\includegraphics{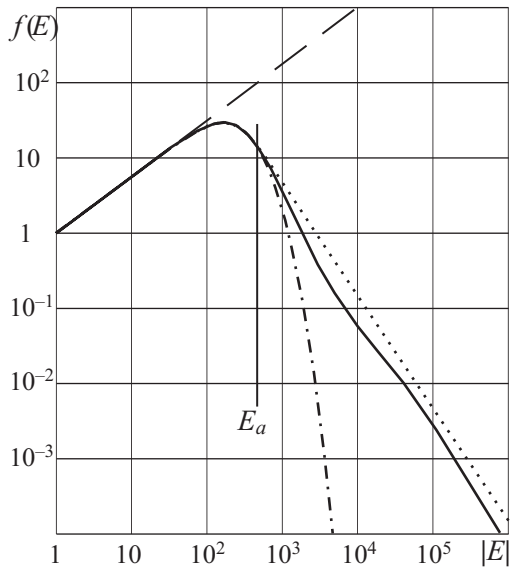} $$
\caption{
$R$-averaged distribution function:
dotted line -- circular orbits ($\beta=-\infty$), solid line -- $\beta=-2$, dotted-dashed line -- $\beta=0$. 
In the latter case $f(E)$ is exponentially small for high $|E|$, while in other cases it tends to 
$|E|^{-3/2}$ corresponding to constant-density core.
Initial distribution function $f(E) \propto |E|^{3/4}$ is shown by long-dashed line. 
The values are normalized to $E_0$ -- energy at the black hole influence radius. Energy $E_a$ corresponding 
to current annihilation radius is denoted by vertical line.
} \label{fig_f_E}
\end{figure} 

\begin{figure}[t] 
$$\includegraphics{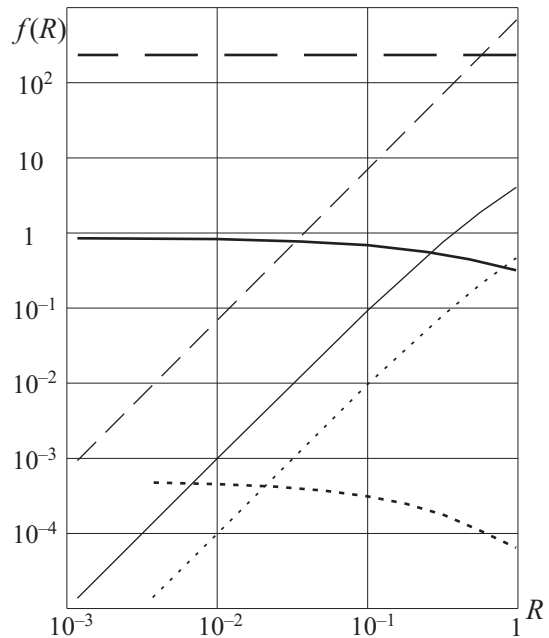} $$
\caption{
$R$-dependent distribution function for certain values of $E$:
long-dashed and solid lines are for $E=3E_a$, $t=0$ and $t=t_1$; 
dotted line is for $E=10E_a$, $t_1$. 
Light (oblique) lines are for $\beta=-2$, heavy (horizontal) lines are for $\beta=0$.
As the values are given for $E>E_a$, the $R$-profile has been changed by annihilation; however, for 
$\beta=-2$ (with density plateau), the annihilation rate is quite independent of $R$ for fixed $E$, 
so the evolution is in the magnitude, not the shape of the distribution function. 
For $\beta=0$, one can see that circular orbits (with $R \sim 1$) are depleted more rapidly than 
radial ($R \sim 0$).
} \label{fig_f_R}
\end{figure} 

The above qualitative arguments have to be confirmed by strict calculations. In order to do this, we solve 
the system of equations (\ref{ann_rate_aver}, \ref{density}) by numerical integration of Eq.(\ref{ann_rate_aver})
forward in time on a rectangular grid in $\{E, R\}$ space, with the density profile recalculated at each time 
step from Eq.(\ref{density}). We start from simple power-law distribution function
\begin{equation}  \label{DF_PL}
f(E,R) = f_0 R^{-\beta} (E/E_0)^{p} \;,\quad c^2/(4GM_{bh})>|E|>|E_0|
\end{equation}
Here the lower boundary $E_0$ defines the energy corresponding to the radius of black hole's influence, 
where the gravitational potential becomes dominated by surrounding stars or dark matter rather than black 
hole itself. 
The higher boundary is determined by black hole horizon. The range of energies under consideration is large, 
say, $10^7$, to avoid boundary effects. The corresponding initial density profile is $\rho_{in}=K\,r^{-(3/2+p)}$.

We have set $p=3/4$ ($\rho \propto r^{-9/4}$), as it corresponds to a NFW halo \cite{a9611107} adiabaticaly 
compressed near a black hole \cite{a9906391}. We scale density and radius to the values at the black hole influence 
radius $r_h$, and consider $r\ll r_h$. The evolution was calculated for $\beta=-2$, i.e. dominance of circular 
orbits, and for $\beta=0$, the case of isotropy. 

Results are shown on Fig.~\ref{fig_dens} for time $t_1$ taken so that annihilation 
plateau density $\rho_a$ equals $10^6$ in scaled dimensionless units [$t_1$ is related to $\rho_a$ by 
Eq.(\ref{dens_evolution_circ})]. The density profile evolves almost self-similarly if we scale $r_a$ and $\rho$ 
simultaneously to remain on initial profile curve, so the value $t_1$ may be taken quite arbitrary.

The results agree well with our preliminary suggestions. 
For the case $\beta<p+1$ ($\beta=-2$ in our calculations) a constant-density plateau develops 
(Fig.~\ref{fig_dens}, solid) with density $\rho_a$ and radius $r_a=(K/\rho_a)^{1/(3/2+p)}$.

In the opposite case a weak cusp $\rho\propto r^{-\gamma_i}, \gamma_i=(1/2+\beta)$ develops, which 
extends up to radius $r_a$ and smoothly joins the initial density profile (Fig.~\ref{fig_dens}, dot-dashed). 
The $R$-averaged distribution function in the latter case is greatly depleted in the region 
$|E|>|E_a|=GM_{bh}/r_a$ compared to the case of density plateau (Fig.~\ref{fig_f_E}). 
So the cusp is indeed formed by particles with rather low energies (high apocentre radii) and high 
eccentricities: if the dominance of radial orbits over circular ones is large enough, then the fraction of 
particles on radial orbits is sufficient to determine the inner density profile.
One can see that the $R$-dependence of distribution function in the case $0<\gamma_i<1$ is biased towards 
radial orbits in the region $|E|>|E_a|$ (Fig.~\ref{fig_f_R}).

We note that the case of isotropic (and even radially anisotropic) velocity distribution is much more 
relevant to cosmological dark matter halos than the case of tangential anisotropy. The parameter $\beta$ in 
the centers of simulated halos is about zero or slightly positive \cite{a0405491}, and in most analytical 
models the situation is the same \cite{a0506528, a0506571}. Indeed a relation between $\gamma$ and $\beta$ 
proposed in \cite{a0411473} suggests $\beta>=-0.15$ for all realistic $\gamma>=0$, which would result in a 
cusp. However, if a binary black hole was present in the center of a halo, it would destroy the cusp and 
generate a core with tangential velocity anisotropy \cite{a0605070}. In this case, however, the annihilation 
plays almost no role because density in the core falls far below $\rho_a$ (the core radius is of order the 
binary separation radius, which is much greater than typical annihilation radius).

\section{Implications for dark matter search} 

The annihilation flux $\Phi$ from the direction of the black hole usually is represented as a product of two 
quantities, the first of them depending on particle physics and the second, called ``astrophysical factor'' 
$\overline J$, is related to dark matter spatial density \cite{h0404175}:
\begin{equation}  \label{J_av}
\overline J = \frac{1}{R_\odot\,\rho_\odot^2} \int_0^{\Theta_{max}} 2\pi\theta\,d\theta 
  \int_{-\infty}^{\infty} dl\, \rho^2(\sqrt{l^2+(R_\odot \theta)^2})\,.
\end{equation}
Here $\Theta_{max}$ is the detector angular resolution (the point-spread function is assumed to be 
of Heaviside form for simplicity), $R_\odot$ is the distance from Sun to the black hole, $\rho_\odot$ is the 
dark matter density near the Sun. The outer integral represents averaging over telescope's angular 
resolution, the inner stands for the line-of-sight integration.  

We may rewrite this expression in terms of the ``vicinity'' of the black hole and the ``background'' from 
outside this vicinity:
\begin{equation}  \label{J_av_2}
\begin{array}{rcl}
\overline J &=& \displaystyle \frac{1}{R_\odot\,\rho_\odot^2} \int_0^{R_{max}} 4\pi r^2\,dr\,\rho(r)^2 + J_{bkg} \;, \\
R_{max} &=& R_\odot \Theta_{max} \,.\mbox{\Large \strut}
\end{array}
\end{equation}
Now we note that for present-day observational capabilities we cannot hope to tell cusp from core, and even 
to resolve the black hole radius of influence.
For example, GLAST will have angular resolution $\Omega_{max}$ of order $0.1^\circ$ \cite{a0612387}, which 
corresponds to spatial distance of 15 pc at the Galactic center ($R_\odot=8.5$ kpc).  
The radius of black hole influence $r_h$ in the center of our Galaxy is about 2 pc, and the annihilation 
radius is even smaller, of order $10^{-3}$ pc \cite{a0504422}.
 
Therefore the integral in (\ref{J_av_2}) is split into three terms: $r<r_a$ is the annihilation plateau or 
inner cusp, $r_a<r<r_h$ is the black hole domain of influence, and $r_h<r<R_{max}$ is the rest. At each of 
these intervals the density is roughly power-law with index $\gamma_k$. It is evident that if $\gamma_k<3/2$, 
then the most part of the integral comes from outer boundary of the corresponding region, and if 
$\gamma_k>3/2$, the integral is determined by inner boundary. 

The density in the case of plateau is given by Eq.(\ref{dens_evolution_circ}), and in the case of weak cusp 
is well approximated by a similar expression:
\begin{equation}  \label{dens_weak_cusp}
\rho(r) \approx \frac{\rho_0(r) \rho_i(r,t)}{\rho_0+\rho_i}\;, \quad
\rho_i = \rho_a(t) \left(\frac{r}{r_a}\right)^{-\gamma_i} \,.
\end{equation}
$\gamma_i$ is the inner cusp power-law index, and is less than 3/2. On the other hand, the outer power-law 
index of the spike is always greater than 3/2 (in our case it equals 9/4). Therefore, the flux from the whole 
black hole domain of influence is determined by $r_a$ only. It appears that the form of transition is 
significant: if we replace (\ref{dens_weak_cusp}) by a simpler formula $\rho(r) = {\rm min} (\rho_0, \rho_i)$,
we overestimate flux almost twice. But the difference between models with $\beta=-2$ (plateau) and $\beta=0$ 
($r^{-1/2}$ cusp) is less than 10\%, which renders the presence of weak cusp almost undetectable. 
Furthermore, the addition of annihilation outside $r_h$, as well as the ``background'', makes the difference 
even smaller. Nevertheless, we want to stress that if the power-law index everywhere outside $r_h$ is greater than 
3/2, then this area contributes little to the total annihilation flux. It is likely to be the case even for 
NFW initial density profile, because it should be adiabatically compressed by baryons during the formation of 
the Galaxy, and the profile becomes steeper than $r^{-3/2}$ \cite{a0601669}.

\section{Conclusion} 

We have reconsidered the problem of dark matter annihilation around a black hole, for the case of arbitrary 
velocity anisotropy of dark matter particles. In the case of circular orbits the result is well-known: 
a constant density core of radius $r_a$ and density $\rho_a$ develops, which smoothly joins the initial 
profile [Eq.(\ref{dens_evolution_circ})]. However, if the fraction of radially biased orbits is large enough, 
that is, if the anisotropy coefficient $\beta$ is greater than $-1/2$, a weak cusp is formed inside $r_a$: 
$\rho \approx \rho_a (r/r_a)^{-(\beta+1/2)}$. The cusp consists of particles which spend most part of their 
orbital period outside $r_a$, but since their orbits are elongated and their fraction is large, they 
contribute enough to the density indide $r_a$. 
The particles with apocentre radii within $r_a$ are annihilated much faster in the latter case, especially on 
orbits with low eccentricities.

However, unless we have a telescope which can resolve the radius $r_a$, we cannot practically distinguish 
between a plateau and a weak cusp.
We note that other dynamical processes, such as scattering of dark matter particles off stars, significantly 
affect dark matter density at radii $r \lesssim r_h$ \cite{a0504422, a0610425, VZnew} 
(They tend to decrease the density if it was sufficiently steep initially). So the effect 
considered in this paper does not seem to play an important role in the evaluation of dark matter 
annihilation signal from the vicinity of black holes. 
It is mostly of terminological significance: the term ``annihilation plateau'' in most cases is misleading, 
and should be replaced by a ``weak cusp''. 

I am grateful to Maxim Zelnikov for helpful discussion, to Steen Hansen for useful remarks, to the referee 
for important comments, and acknowledge support from Russian Fund for Basic Research (grant No. 07-02-01128-a) 
and from President's grant council (grant No. NSh-4407.2006.2).

\end{document}